# Title

Generalised Time-Series Analysis of Fault Mechanics Using Explainable AI

## Authors

Thomas King[1,2], Sergio Vinciguerra[2*]

## Affiliations

[1] Birmingham Institute for Sustainability and Climate Action, University of Birmingham, Edgbaston, Birmingham, B15 2TT, United Kingdom.

[2*] Department of Earth Sciences, University of Turin, Via Verdi 8 – 10124, Torino, Italy. Email: sergiocarmelo.vinciguerra@unito.it.

## Abstract

Understanding how faults nucleate and grow is a critical problem in earthquake science and hazard assessment. This study examines fault development in Alzo granite under triaxial pressures ranging from 5 to 40 MPa by applying a Time Delay Neural Network (TDNN) to multi-parameter acoustic emission (AE) data. The TDNN integrates waveform-derived attributes, including peak delay and scattering attenuation, with occurrence-based metrics such as time distributions, Gutenberg-Richter b-values, and spatial fractal dimensions, to characterize the transition from distributed microcracking to localised faulting. Optimised via genetic algorithms, the TDNN dynamically weights these parameters, enabling accurate characterisation of fault growth stages. Our results delineate three distinct phases of fault evolution: nucleation of random microcracks indicated by changes in elastic wave scattering, initiation of fault growth reflected in evolving AE spatial and magnitude distributions, and fault coalescence marked by exponential increases in peak delay and b-value shifts. The model predicts the timing and magnitude of stress drops across varying pressures and failure mechanisms, from axial splitting to shear localisation, providing deeper insights into fault mechanics through explainable AI models.

## Introduction

Understanding fault mechanics and nucleation remains crucial for earthquake science and hazard assessment. The failure of intact rocks in the brittle regime and the mechanisms of fault nucleation and propagation have been extensively studied since the early 1990s. These studies draw parallels between rupture nucleation under laboratory triaxial conditions and natural slip nucleation in earthquakes (1–3). Acoustic emissions (AE), elastic waves generated by rapid cracking in solids, have become vital for tracking the progression of damage before failure, as well as for monitoring fault nucleation and growth. AE waveform attributes, including frequency content, amplitude, and duration, change dynamically depending on load, deformation conditions, and material properties. The objectives of AE studies typically include (i) characterizing individual AE events to link them to micro-mechanisms (5), (ii) applying statistical relationships such as the Gutenberg-Richter law and fractal clustering to better understand deformation rates and processes (6, 7), and (iii) localizing AE sources to determine whether deformation is distributed or localised (8, 9).

Fault development under triaxial compression follows distinct stages: (i) early nucleation of random tensile microcracks with limited interaction, (ii) the formation of a process zone as microcracks begin to interact near the rock's ultimate strength, and (iii) propagation of this process zone, leading to the formation of a fault nucleus and continued microcrack dilation (1). Recent research highlights the importance of dilatancy and compaction patterns, with advanced techniques such as X-ray tomography providing deeper insights into these processes (10). Studies have also shown that deformation mechanisms transition from axial splitting at lower confining pressures to shear localization at higher pressures, as observed in experiments on Westerly granite (11). AE focal mechanism analyses indicate that deformation is often dominated by compactant (C-type) and dilatant (T-type) regions. C-type events represent the nucleation of early microcracks, while T-type events signal the cycles of new crack formation and coalescence (9).

Waveform-derived attributes, such as peak delays, coda Q attenuation, and energy diffusion length scales, have proven essential in mapping variations in subsurface heterogeneity. These attributes have been extensively used in natural earthquake studies to reveal differences in material properties, fracture density, and fluid saturation (12). Peak delays, for example, have helped identify thrust faults and sedimentary structures, offering insights into the control mechanisms behind fluid overpressure (12). Similarly, coda Q attenuation measures energy dissipation mechanisms such as crack closure and pore fluid movement, providing valuable information on evolving strain conditions in tectonically active regions (13, 14). These methodologies are applicable to AE data but require interpretive models to link waveform metrics to underlying deformation processes, as variations in attributes like coda Q attenuation do not always correspond directly to changes in damage zones or crack density (15–19).

Despite significant advances, predicting rupture mechanisms remains a challenge due to rock heterogeneity and the complex evolution of strain under varying pressure conditions. Probabilistic methods and machine learning (ML) offer promising solutions for addressing these challenges. ML techniques, which have a long history in seismology, have expanded rapidly with improvements in computational power and neural networks (20). ML is now employed in earthquake detection, phase picking, and seismogram simulation (21). Recent studies have applied ML models to predict time-to-failure (TTF) using AE data from double-shear experiments, allowing for forecasts of critical parameters such as shear stress, fault friction, and slip velocity (19, 22–24). More sophisticated models, including transformer networks, have shown success in predicting stress variations in laboratory fault zones (25). However, many models still lack full integration of evolving spatial and seismic attributes. Methods incorporating spatial clustering features, such as fractal dimension analysis, have demonstrated potential for improving event forecasting by combining spatial and temporal information (18, 26–28).

This study focuses on investigating the fracturing behaviour of Alzo granite (NW Italy) through a combination of laboratory experiments and Time Delay Neural Networks (TDNN). Samples of granite were subjected to triaxial compression tests at pressures of 5, 10, 20, and 40 MPa to induce deformation and fracturing. AE data were collected and analysed to extract key parameters, including peak delay, diffusion properties, AE event rate, fractal dimension, and Gutenberg-Richter b-values. These parameters provide critical insights into the evolution of fractures and fault structures. Hyperparameters for the TDNN were optimised using a genetic algorithm (GA), allowing the network to classify time-dependent trends in the AE datasets as they correlated with changes in stress and strain. By combining physical experimentation with neural network optimization, this study aims to enhance the understanding of fracturing processes in heterogeneous rock formations.

**Results**

*Training Data*

Training data for the TDNN were gathered from triaxial deformation experiments at confining pressures of 5, 10, 20, and 40 MPa, covering a wide range of failure mechanisms. Lower pressures led to axial splitting, characterised by vertical fractures, while higher pressures induced planar shear failure (9). AE-derived parameters were smoothed using a moving window approach (20% UCS), with time-dependent variations normalised to strain at failure. To account for differences in sample behaviour post-failure, strain shifts were applied to align data across different experiments. This normalization was essential to identify consistent trends in the evolving AE signals, and the average behaviour of each parameter is depicted in Figure 1.

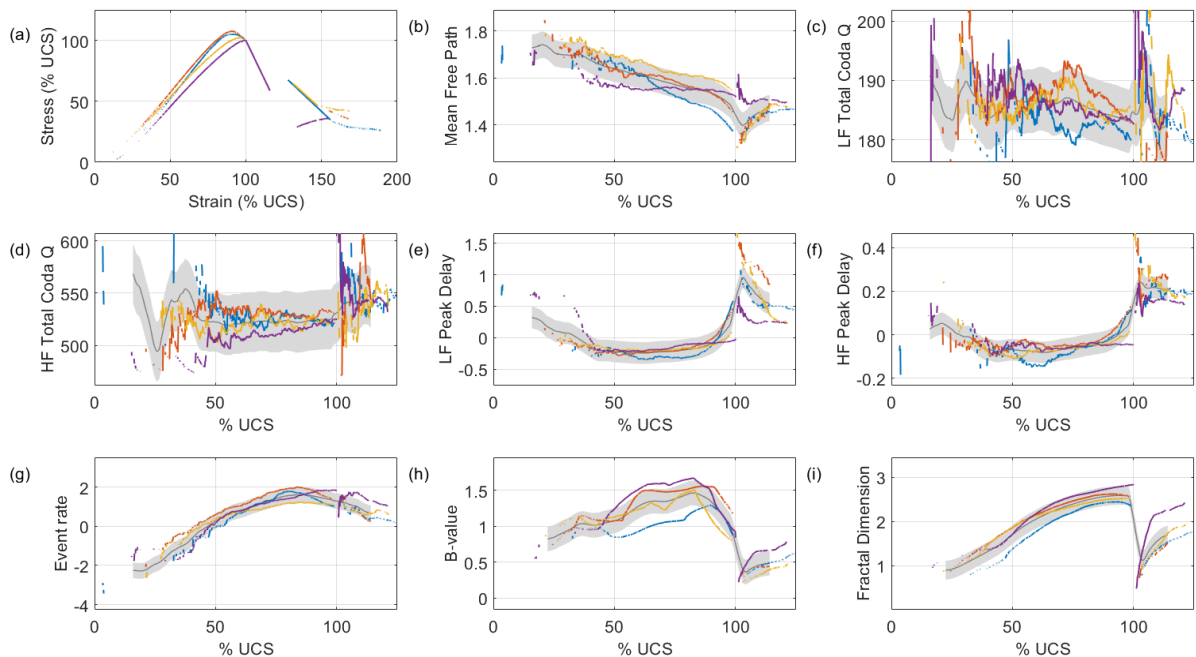

Figure 1: Training data for the neural network to learn taken from conventional triaxial deformation experiments conducted at 5 MPa (blue), 10 MPa (orange), 20 MPa (yellow) and 40 MPa (purple). Average variations for all experiments are shown with a black line. a) Stress-strain mechanical data recorded during the experiments are normalised against their values at sample failure (i.e., 100% UCS) and are used as the target classification in separate neural networks. For training purpose, post-failure variations are time-shifted removing the strain change failure. Missing data indicate periods where no AE activity was detected. Trends for model input training parameters are calculated and smoothed in a moving window using the most recent 20% Ultimate compressive strength (UCS) of data. Time-dependent variations are normalised to strain value at failure and presented as a percentage of UCS. Waveform derived: b) Mean free path (λ) derived from diffusion coefficient estimates. c) Low frequency total coda Q (LF Q). d) High frequency total coda Q (HF Q). e) Low frequency relative peak delay (LF ΔPD). f) High frequency

relative peak delay (HF ΔPD). Occurrence derived: g) Inverse logarithm of inter-event timings (Event rate). h) B-value estimates calculated from estimated signal amplitudes. i) Fractal dimension estimates calculated from event source locations.

Among waveform-derived parameters, mean free path (λ, Figure 1b) and peak delay (PD, Figures 1e–f) showed the strongest sensitivity to experimental conditions. λ exhibited a linear decline from 1.7 cm at early loading to approximately 1.4 cm at sample failure (100% UCS). A rapid decrease occurred near 95% UCS, indicating increased scattering of wave energy as fractures approached coalescence. Interestingly, a minor rebound was observed in λ post-failure, potentially linked to structural rearrangements within the fault zone. Peak delay (PD) values decreased initially but began rising exponentially near 90% UCS, peaking at sample failure. This behaviour, observed across both low-frequency (LF) and high-frequency (HF) components, highlights the increased complexity of wave propagation through fractured material. HF PD values showed a more stable peak response post-failure compared to LF PD. Estimates for total coda Q (Figures 1c–d) remained relatively constant during early loading stages, with no significant differences across confining pressures. However, a subtle increase in HF coda Q was noted at failure, potentially linked to fault coalescence.

Occurrence-derived parameters displayed even stronger correlations across experimental conditions. AE event rate (e, Figure 1g) increased steadily during early loading, flattening near 80% UCS before declining post-failure. This flattening coincides with the formation of interconnected fracture networks. Both b-value (Figure 1h) and fractal dimension (Figure 1i) followed similar trends, peaking around 80–100% UCS. B-value demonstrated greater variability across pressures, with additional peaks at ~40% and ~60% UCS. These peaks may reflect intermediate stages of fracture interaction before full coalescence. The fractal dimension's consistent increase to a peak at failure emphasises its role in capturing spatial clustering within evolving fault structures.

*Neural Network Classification*

Observed (black) and average predicted (red) variations in normalised strain are presented in Figure 2. For each confining pressure, the model was trained on data from three other tests and validated against the presented condition. Error bars show TDNN output variability across 10 training runs. Early experimental stages (<30% UCS) had limited characterization due to sparse AE data. At 5 MPa (Figure 2a), predictions closely followed observed data, with maximum pre-failure errors of 5% UCS. However, post-failure trends were less accurately captured. At 10 MPa (Figure 2b), alignment improved pre-failure, though strain changes at failure were inadequately modelled. For 20 MPa (Figure 2c), the model exhibited good accuracy both pre- and post-failure. Predictions at 40 MPa (Figure 2d) showed uncertainty during early stages, though fault growth and strain change at failure were well captured.

Differential stress predictions (Figure 3) presented more variability due to complex non-linear behaviour. Stress hardening during early loading (<50% UCS) transitioned to strain softening during later stages. Although trends were broadly modelled, subtle phase transitions were less accurately represented. Peak stress timing was well captured at 5 and 10 MPa but exhibited errors of up to 10% UCS at higher pressures. Despite these challenges, the model predicted stress drops at failure with reasonable accuracy, though hyperparameter variation led to wider error margins.

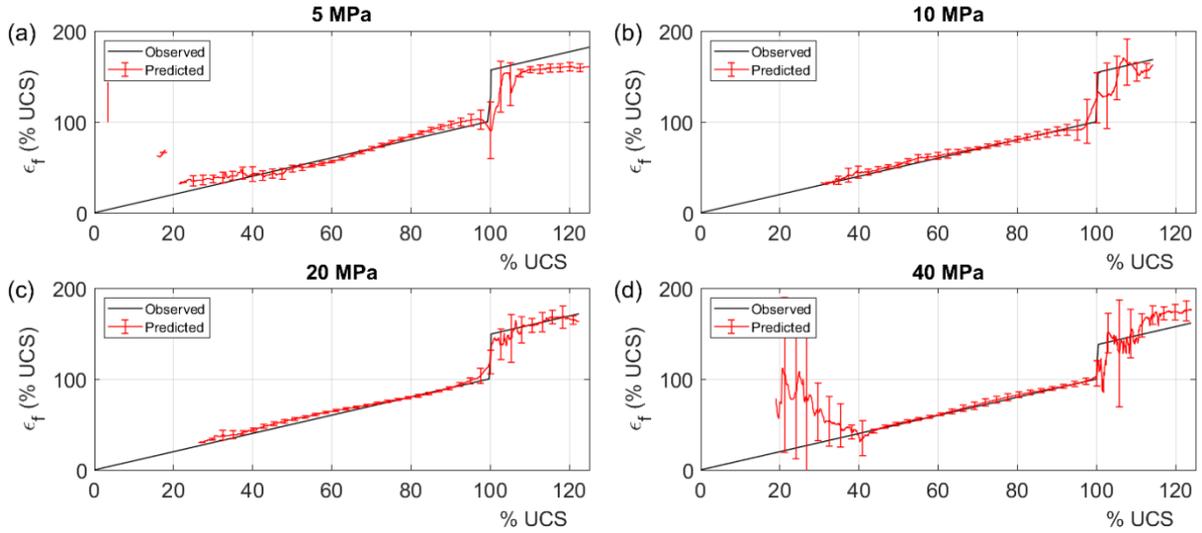

Figure 2: Observed strain variations (black) normalised to strain at failure ($\varepsilon_f$) and plotted as a percentage of ultimate compressive strength (UCS). Average predicted strain variations (red) for 10 successive runs of the neural network training routine. Error bars are calculated as the standard deviation of all runs. Predicted variations are calculated using data from three experimental conditions and validated on presented condition: a) 5 MPa, b) 10 MPa, c) 20 MPa, d) 40 MPa.

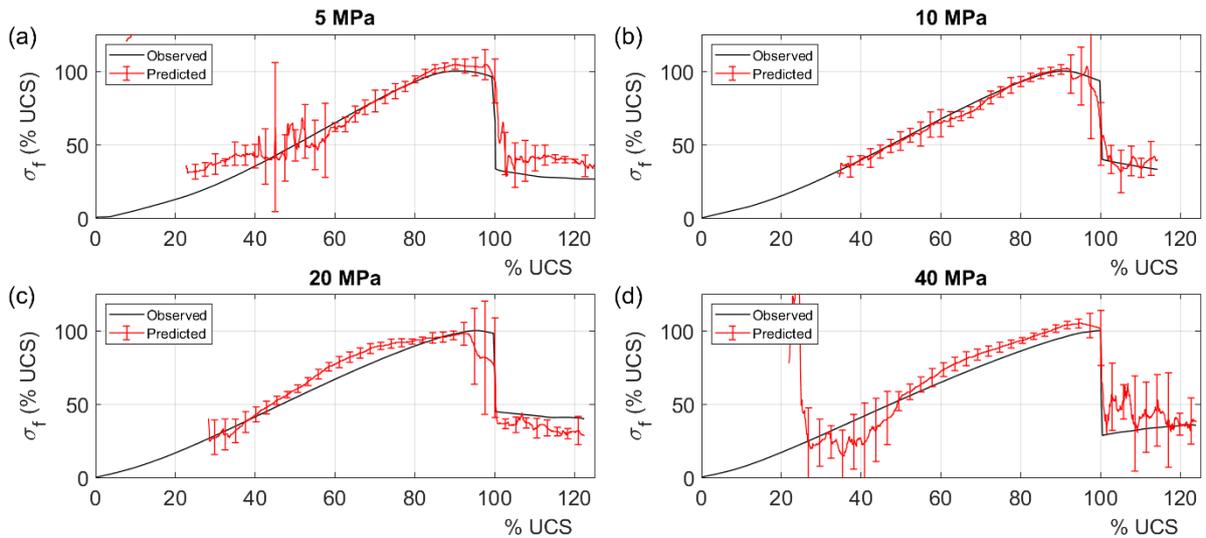

Figure 3: Observed differential stress variations (black) normalised to stress at ultimate compressive strength (UCS, $\sigma_f$) and plotted as a percentage of UCS. Average predicted differential stress variations (red) for 10 successive runs of the neural network training routine. Error bars are calculated as the standard deviation of all runs. Predicted variations are calculated using data from three experimental conditions and validated on presented condition: a) 5 MPa, b) 10 MPa, c) 20 MPa, d) 40 MPa.

*Neural Network Parameter Sensitivity*

The TDNN training process assigns initial random weights to each input parameter, adjusting them over iterations to optimise classification accuracy. To evaluate time-dependent parameter importance, models were trained using a moving window approach, with each window spanning 20% UCS and shifting in 5% increments. Figure 4 illustrates relative parameter importance for both strain (Figure 4a) and stress (Figure 4b) predictions. In the early stages of fracture nucleation (<30% UCS), waveform-derived parameters (WD) held greater significance. However, as fracture growth progressed (30–60% UCS), occurrence-derived parameters (OD) gained prominence. Near fault coalescence (~90% UCS), WD parameters exhibited a surge in importance, which remained elevated post-failure.

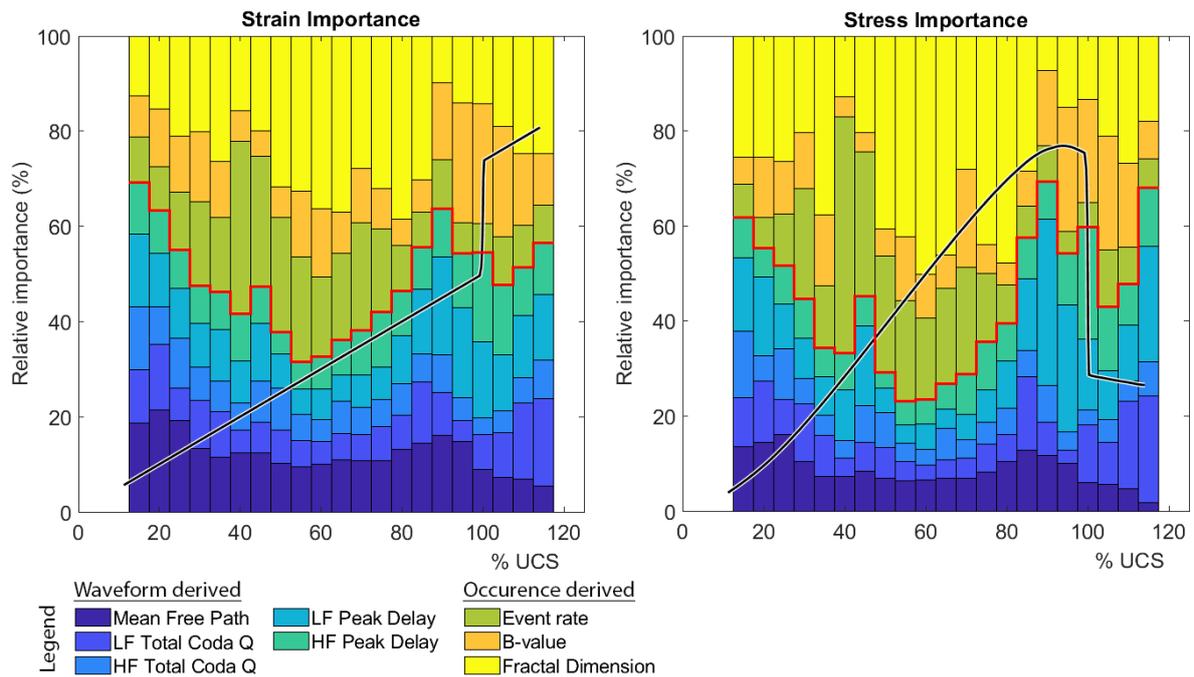

Figure 4: Relative parameter importance variations derived from assigned neural network weights calculated in a moving window of 20% ultimate compressive strength (UCS) for strain (left) and stress (right) variations (black lines). A red line is used to separate contributions of waveform and occurrence derived training data. Low frequency (LF) and high frequency (HF) components for waveform-derived parameters are separately presented.

Among individual parameters, mean free path maintained consistent importance (~10–20%) across both models. Total coda Q and peak delay demonstrated exponential importance increases between 60% and 80% UCS. Low-frequency (LF) total coda Q peaked at ~85% UCS, preceding a sharp rise in LF peak delay near 90% UCS. AE event rate rose significantly at 40% UCS but declined as fault networks became established. The b-value showed relatively low importance during early stages but rose to ~25% at fault coalescence. Spatial fractal dimension consistently ranked as the most critical parameter, maintaining an importance level of 25–50% during much of the fault evolution process.

**Discussion**

Many aspects of fault development exhibit scale-invariant behaviour, meaning the underlying mechanisms are consistent across both laboratory and geological scales. This consistency suggests that microcrack nucleation, interaction, and fault coalescence in small samples can provide insights into large-scale fault systems. However, significant challenges arise due to material heterogeneity, differences in fault dimensions, and variations in timescales. These factors complicate the development of generalised forecasting models. Despite these difficulties, the results of this study demonstrate that Time Delay Neural Networks (TDNN) can effectively model the progression of damage and fault growth through a dynamic weighting of multi-parameter AE data.

In the early deformation stages (<40% UCS), the TDNN assigned high importance to waveform-derived parameters, such as peak delay and mean free path. These parameters are sensitive to initial microcrack nucleation, reflecting increased heterogeneity in wave propagation paths as cracks develop. Similar findings have been reported in studies where waveform scattering is linked to microfracture distributions (32, 33). As deformation progressed (40–80% UCS), the model shifted to prioritise occurrence-derived parameters, such as fractal dimension and AE event rate. This shift aligns with the formation of localised damage zones and the clustering of fractures. Near sample failure (>80% UCS), the TDNN returned to emphasizing waveform parameters, including peak delay and coda Q attenuation. The increase in peak delay, observed as failure approached, is consistent with heightened wave scattering along newly coalescing fault networks (14, 15).

The model also highlighted the role of the b-value, which increased sharply at fault coalescence. In both laboratory and field studies, a declining b-value often signals the formation of larger fractures, indicating imminent failure. This study's results confirm that the TDNN captured this critical transition, with b-value importance peaking at ~90% UCS. Spatial fractal dimension, which maintained high importance throughout, underscores the importance of monitoring the spatial distribution of evolving cracks.

These findings demonstrate that TDNNs can bridge the gap between laboratory-scale observations and larger geological processes by dynamically integrating waveform and occurrence metrics. The model successfully captured the transition from distributed microcracking to localised faulting across various confining pressures and failure modes. This adaptability highlights the model's potential applicability to diverse geological settings and fault mechanisms.

Scaling laboratory results to field conditions remains challenging due to increased structural complexity and the influence of fluids on fault networks. While laboratory AE data offer detailed, controlled observations, real-world fault systems present additional uncertainties. Nonetheless, machine learning models such as TDNNs show promise for enhancing real-time monitoring and early warning systems. By focusing on key parameters like peak delay, coda Q, and fractal dimension, these models can identify critical stages of fault development in natural settings.

Future research should explore variations in rock lithology and potential dependencies on scale to generalise these findings. Additionally, integrating spatial clustering features with continuous AE monitoring may improve the detection of early fault nucleation and growth. The combination of laboratory experiments and explainable AI approaches, as demonstrated here, provides a pathway toward more effective monitoring and risk assessment for evolving fault systems.

This study demonstrates that Time Delay Neural Networks, optimised through genetic algorithms, provide critical insights into fault mechanics by dynamically prioritizing key AE parameters. By integrating both waveform-derived and occurrence-based metrics, the model successfully characterised fault nucleation and growth stages, capturing transitions from microcracking to localised faulting. These findings contribute to advancing real-time monitoring capabilities for natural and engineered fault systems, with potential applications in seismic hazard assessment and infrastructure stability monitoring.

## Materials and Methods

*Sample Preparation and Experimental Setup*

Alzo granite is a fine to medium grained igneous rock sourced from the Piedmonte region in Northern Italy. It consists of interlocking crystals of quartz, feldspar and biotite that result in a porosity value of 0.72±0.1% and an approximate density of 2700 kg/m$^3$ (*36*). Four 40x100mm samples were prepared from pristine quarry samples with a diamond tipped hollow coring drill. End faces were then trimmed to length and ground flat to within 0.01mm with a cross-cutting diamond disc. Samples were positioned within an engineered rubber jacket instrumented with an array of twelve evenly distributed 1 MHz Piezo-Electric Transducers (AE sensors, PAC Nano30) that are used to detect the elastic wave energy produced by micro-fracturing events. Fracturing of the samples is induced via conventional triaxial cell installed at the University of Portsmouth, UK (e.g., *37*). A high flash-point oil is used as a confining medium and provides confining pressure along the radial boundaries of the samples. Axial load is applied from above with a vertical piston at a constant strain rate of 3.6mm/hr measured by an externally mounted non-contact eddy displacement system. This enables direct measurement of differential stress and sample strain as each experiment progresses. For a more detailed overview of these laboratory experiments and the AE data please refer to (*9*).

*Acoustic Emission Attributes – Waveform Derived*

The time delay between signal onset and the maximum amplitude arrival is conventionally known as the peak delay (*32*). The occurrence of high peak delays in AE is generally attributed to the presence of heterogeneity along the ray path resulting in trapped waves propagating along faulted structure to nearby sensors (*15, 16*). AE waveforms are first filtered within the low and high frequency windows and the peak delay estimated from the root mean square envelope of the filtered data. Measurements are then averaged over the 12 waveforms detected for each event. Hereafter, peak delay is considered as the relative value $\Delta \log(PD_f)$ that varies around the average of all (*N* number) measurements for each frequency band (*f*):

$$\Delta \log(PD_f) = \log(PD_f) - \frac{1}{n}\sum_{i=1}^{N} \log(PD_f).$$

The diffusion approximation in seismology provides a framework to model the propagation of energy within very heterogeneous media (*33*). This method assumes an isotropic scattering environment, treating waveform propagation in a manner similar to the diffusion of particles or heat. This scattering is modelled by equation 1 that describes the evolution of seismic energy density *E(r,t)* over time *t* and space *r*:

$$\frac{\Delta E(r,t)}{\Delta t} = d\nabla^2 E(r,t) - \frac{E(r,t)}{\tau},$$

where *d* is the diffusion coefficient, $\tau$ is the characteristic absorption time and $\nabla^2$ is the Laplacian operator. The solution to the diffusion equation for a point source with initial energy $E_0$ and damping coefficient $\beta$ is:

$$E(r,t) = \frac{E_0}{(4\pi dt)^{3/2}} \exp\left(-\frac{r^2}{4dt} - \beta t\right).$$

For each of the detected AE, waveforms are first bandpass filtered within the low and high frequency windows using a 2nd order Butterworth filter. The left side of the diffusion equation, *E(r,t)*, is then estimated as the average waveform energy within a moving window of 0.1 ms calculated from signal onset to 1 ms after the onset. To help stabilise solutions for *d, E₀* and *β* obtained through a linear least squares inversion, all 12 waveforms for a single AE event are included in the model vector.

To better relate solved coefficients *d* and *β* to the physical medium and provide more direct estimates of intrinsic and scattering attenuation, we use the total coda quality factor Q and the mean free path $\ell$. Under the diffusion approximation, the relationship between Q and the damping coefficient *β* is given as:

$$Q = \pi f / \beta.$$

The mean free path, $\ell$, represents the average distance elastic wave energy travels before scattering. It is derived from the diffusion coefficient at frequency, *f*, through the following equation:

$$\ell = \sqrt{d f^{-1} / \pi}.$$

*Acoustic Emission Attributes – Occurrence Derived*

The AE event rate, *e*, provides an intuitive measure of how many fracturing events are occurring within a specific timeframe. It has been previously identified to be a key parameter in strategies for forecasting of failure in material sciences (*38*), volcanic eruptions (*39*), laboratory biaxial experiments (*18*, *26*) and fracking operations (*40*). For this study it is defined as:

$$e = \ln(\Delta T),$$

where $\Delta T$ is the relative time difference between successive AE irrespective of source location.

Fractal dimension quantifies the complexity in the spatial distribution of fracturing events and is commonly used to provide insight into the development of fault structure (*34*), as it measures the degree of AE spatial clustering (*35*). For this study it is calculated using the box-counting method within a moving window of 20% UCS to quantify the different stages of fault evolution (*41*). The medium is discretised into a 3D grid and for each AE source location, the number of grid points that falls within a minimum distance to the event is counted. The following ratio then provides an estimate for fractal dimension, $D_F$:

$$D_F = \frac{\log(count)}{\log(grid\ size)}.$$

Derived from the Gutenberg-Richter law, the b-value represents the slope of the frequency-magnitude distribution of fracturing events and is expressed as:

$$log N(M) = a - bM,$$

where *N(M)* is the number of events with magnitude greater than or equal to M, *a* is a constant that represents the total seismicity rate and *b* indicates the relative likelihood of a large or small event, thus providing key info on the source size time evolution during the test. To analyse the temporal variation in AE magnitudes, data are windowed in a moving window of 20% UCS. The magnitude range for analysis is defined as the 5$^{th}$ and 95$^{th}$ percentile of all event magnitudes to focus on the central distribution. The b-value is calculated here by counting the number of events that exceed a specific magnitude, *N(M)*, and solving a linear system of equations to obtain *a* and *b* that best fit the observed count (*42*).

*Neural Network Optimisation*

Time Delay Neural Networks (TDNN) are a class of artificial neural networks specifically designed to classify sequential patterns within timeseries data (*29*). Analogous to convolutional neural networks, input series of data are passed through convolution filters that slide over the temporal dimension. Each cluster in a time-delay layer is connected to a subset of input neurons that correspond to a particular time interval, allowing for the classification of shift-invariant trends in the data. This architecture allows the network to learn from previous timesteps and to recognise patterns that are not immediately adjacent. During training, TDNN assigns random initial weights to the input parameters. Although these are adjusted during each iteration of the training routine, the model can converge to different solutions that prioritise different temporal aspects of the data. Moreover, the performance of a TDNN and its sensitivity to a particular training parameter (e.g., AE event rate) are heavily influenced by its hyperparameter architecture, the number of hidden layers, the number of neurons in each layer, and the time-delays for each layer. This aspect is highly non-linear with a strong interdependence between the hyperparameters that describe the model architecture and the resulting sensitivity.

Genetic algorithms (GA) are a global optimisation technique that use evolutionary-inspired mechanisms to more robustly explore the model search space compared to gradient-based methods (*16*). It is used here to optimise the hidden layer architecture of the TDNN in classifying time-dependent trends in the AE derived datasets as corresponding values for stress and strain. The GA iteratively evolved a population of candidate architectures utilising the Levenberg-Marquardt algorithm, evaluating each using a custom objective function that trained and validated the TDNN on cross-validated datasets. The objective function calculated the mean squared error between the model output and target classification and summed these values to derive the fitness score. Target values for model misfit (10% error) and maximum training time (30 seconds) were set as convergence goals to limit computational overheads. These were defined through qualitative estimates of model accuracy during development. Although, this approach may limit more accurate solutions it was considered sufficient for this phase of the training routine as it helped to avoid overfitting of the data. Once optimal hyperparameters are obtained, defined when population diversity reached a minimum threshold, new TDNN are trained with that architecture for a longer time (5 minutes) and with stricter requirements for model misfit (1% error) to obtain the final results.

**Acknowledgments**

The authors would like to thank numerous colleagues and members of the community for their valuable comments and feedback during the development of this research.

**Funding**

The first author would like to acknowledge the support from the DAFNI-ROSE grant (EP/V054082/1) that has provided access to the local HPC in the National Buried Infrastructure Facility (NBIF).

**Author contributions**

Conceptualization: TK, SV

Methodology: TK

Investigation: TK, SV

Visualization: TK

Supervision: SV

Writing—original draft: TK, SV

Writing—review & editing: TK, SV

**Competing interests**

All other authors declare they have no competing interests.

**Data and materials availability**

The data and code used in these analyses are currently in a developmental format and may require additional guidance for use. Researchers interested in reproducing or extending these analyses are encouraged to contact the authors directly for assistance. The authors are committed to sharing the data and code but note that successful utilisation will likely require coordinated collaboration to ensure proper understanding and application.